# Experimental studies of the fractional quantum Hall effect in the first excited Landau level

W. Pan
Sandia National Laboratories

J.S. Xia
University of Florida and National High Magnetic Field Laboratory

H.L. Stormer
Columbia University and Bell Labs, Alcatel-Lucent Inc.

D.C. Tsui
Princeton University

C. Vicente, E.D. Adams, and N.S. Sullivan
University of Florida and National High Magnetic Field Laboratory

L.N. Pfeiffer, K.W. Baldwin, and K.W. West
Bell Labs, Alcatel-Lucent Inc.

Abstract

We present a spectrum of experimental data on the fractional quantum Hall effect (FQHE) states in the first excited Landau level, obtained in an ultrahigh mobility two-dimensional electron system (2DES) and at very low temperatures and report the following results: For the even-denominator FQHE states, the sample dependence of the $\nu=5/2$ state clearly shows that disorder plays an important role in determining the energy gap at $\nu=5/2$. For the developing $\nu=19/8$ FQHE state the temperature dependence of the $R_{xx}$ minimum implies an energy gap of ~5mK. The energy gaps of the odd-denominator FQHE states at $\nu=7/3$ and $8/3$ also increase with decreasing disorder, similar to the gap at $5/2$ state. Unexpectedly and contrary to earlier data on lower mobility samples, in this ultra-high quality specimen, the $\nu=13/5$ state is missing, while its particle-hole conjugate state, the $\nu=12/5$ state, is a fully developed FQHE state. We speculate that this disappearance might indicate a spin polarization of the $\nu=13/5$ state. Finally, the temperature dependence is studied for the two-reentrant integer quantum Hall states around $\nu=5/2$ and is found to show a very narrow temperature range for the transition from quantized to classical value.

Introduction

Since the discovery of the fractional quantum Hall effect (FQHE) state at Landau level filling $\nu=1/3$ [1,2], many FQHE states have been discovered [3-5] in the lowest (N=0) Landau level. In Table I, we have listed, to our best knowledge, all odd-denominator FQHE states that have been identified in this Landau level (black and gray font). Remarkably, almost all these FQHE states, more than 90% (black font), can be mapped onto an integer quantum Hall effect (IQHE) state of composite fermions (CFs) [6-12]. The remaining fractions (gray font) which cannot be mapped onto IQHE states of CFs, are viewed as FQHE states of CFs [13], demonstrating the importance of residual interaction between CFs.

No FQHE states have been observed in high Landau levels (N≥2). The additional nodes in the electron wavefunction in these Landau levels effectively suppresses the short range electron-electron interaction and, as a result, unidirectional electron density wave state (also called "stripe phases") and Wigner-solid states of electron clusters (also termed "bubble phases") win out over the FQHE states as the ground state. When the electron temperature is raised, a melting transition from the correlated electron solid to a correlated electron liquid is observed and evidence of FQHE states is seen at two very high Landau fillings, $\nu=21/5$ and $24/5$ (underlined italic black font) [14].

In the first excited (N=1) Landau level, which is the focus of this paper, the FQHE has been observed at even-denominators $\nu=5/2$, $7/2$, and $19/8$ (italic gray font), as well as at several odd-denominator fillings (italic black font) [15-21]. Compared to the N=0 Landau level, the FQHE states in the N=1 Landau level are quite unusual. Most of them cannot be viewed as the IQHE states of CFs. The most bizarre among them and the most studied is the state at $\nu=5/2$ [22]. This state does not follow the odd-denominator rule set by the initial Laughlin wavefunction, and today is believed to be due to paring of CFs [23-31]. In loose analogy to the formation of Cooper pairs in superconductivity [32], this pairing of CFs creates a gapped, BCS-like ground state at $\nu=5/2$, which displays the FQHE.

Besides the even-denominator FQHE states, the odd-denominator FQHE states of the N=1 Landau level are of interest as well. As compared to the lowest Landau level, fewer odd-denominator FQHE states are observed in the N=1 Landau level, and their physical origin is yet to be firmly established. For all these reasons, the physics underlying the N=1 FQHE states remains of great interest [33-39]. In particular, the proposal of using the conjectured non-abelian quasi-particles of the $\nu$=5/2 FQHE states for topologically protected, fault tolerant quantum computation has created considerable excitement [40,41]. On the experimental side, results on the FQHE states in the N=1 Landau level have been scarce, due to the extraordinary requirements on high sample quality and low electron temperature. In this paper, we present recent data, obtained in a specimen of ultra-high electron mobility and recorded at very low temperature, that provide new insight into the properties of the correlated states in the N=1 Landau level.

The paper is structured as follows: Section 2 details the sample parameters and experimental techniques. Section 3 presents the main experimental results and discussions. Summaries and the discussion of open issues are provided in Section 4.

2. Sample and experimental techniques

The specimen is a quantum well, symmetrically doped on both sides at a setback distance of 100nm. The well width is 30 nm. The electron density, n=3.1x10$^{11}$ cm$^{-2}$ and mobility, $\mu$=31x10$^6$ cm$^2$/Vs were established after illuminating the specimen with a red light emitting diode (LED) at low temperature (*T*). The two-dimensional electron density differed by 1-5% from one cool-down to another. Within any given cool-down, the electron density stabilized only after being kept cold (*T* < 0.3K) for ~24 hours. All data were taken after this interval.

Ultra-low *T* measurements were carried out in the same demagnetization refrigerator as in Ref. [16]. Specially designed sintered silver heat exchangers were used to cool the two-dimensional electron system (2DES). The fridge temperature was monitored by a CMN thermometer, a $^3$He melting curve thermometer, and a Pt-NMR thermometer. All measurements were performed in an ultra quiet environment, shielded from electro-magnetic noise. Standard low-frequency technique (~ 7Hz) was utilized to measure the magnetoresistance $R_{xx}$ and the Hall resistance $R_{xy}$, with an excitation current of 1nA.

During the course of the experiments, we found the sample state to be very sensitive to its cooling and illumination history. Several different illumination protocols were tested, such as illuminating the sample continuously from room temperature and stopping at 10K, 4.2K, 1.2K, or cooling the sample in the dark and then illuminating it at 10K, 4.2K, 1.2K for 30 minutes. The cleanest $R_{xx}$ and $R_{xy}$ features were obtained by cooling the sample in the dark and then illuminating it at 4.2 K for 30 minutes, at an LED current of 30 $\mu$A.

3. Experimental results and discussions

3.1   Disorder in the $\nu$=5/2 FQHE state

The first even-denominator FQHE state at $\nu$=5/2, discovered in 1987 [15] and unequivocally demonstrated to be quantized in 1999 [16] remains enigmatic. Not following the initial "odd-denominator" rule of the lowest Landau level, its underlying physics has been hotly debated. At present, theory seems to gravitate towards it being a condensed state of CF pairs [22-31] with quasi-particle excitations of non-abelian statistics [41], the so called "Pfaffian state". This has led to an emerging effort to exploit this system for quantum computation [40,41]. However, on the experimental side, neither CF pairing, nor the bizarre statistics of its quasi-particles has been demonstrated yet.

At this stage we only have comparisons between measured and calculated energy gaps to support (or reject) the notion of a paired CF ground state at $\nu$=5/2. All previous data show an energy gap [16-18,20,42] that is much smaller than the theoretically value [26,27]. For example, a 2DES of mobility $\mu$=17x10$^6$ cm$^2$/Vs showed an energy gap of ~ 0.1K [16] which is over one order of magnitude smaller than the theoretical value [26,27]. In order to reconcile this difference, an *ad hoc* disorder broadening of ~2 K must be assumed [16], which, taking up 95% of the gap, is rather unphysical and exceeds a broadening estimated from the mobility by a factor of 300. Thus, the role of disorder in determining the size of the many-body energy gap at 5/2, or in general, the stability of the 5/2 state, remains to be understood.

In the present high quality sample, the $\nu=5/2$ state is particularly strong. In fact, $R_{xx}$ remains vanishingly small even at a temperature of ~ 50mK. An energy gap of $\Delta_{5/2}$ ~ 0.45K is deduced from the Arrhenius plot of Fig. 1a, using $R_{xx} \propto \exp(-\Delta_{5/2}/2K_BT)$. This large value in a high quality specimen emphasizes the importance of residual disorder to determining the size of the $\nu=5/2$ gap.

To quantify the role of disorder, we have measured the energy gap at $\nu=5/2$ in a series of high quality samples. Table II lists the sample parameters. Fig. 1b shows the energy gap $\Delta^{norm}=\Delta_{5/2}/e^2/\varepsilon l_B$, normalized to the strengths of the electron-electron interaction $e^2/\varepsilon l_B$, as a function of $1/\mu$, which is proportional $1/\tau$ and hence proportional to a lifetime broadening. In these equations e is the electron charge, $\varepsilon=12.8$ is the dielectric constant of GaAs, and $l_B = (\hbar/eB)^{1/2}$ is the magnetic length. This plot also includes results obtained by Eisenstein et al [17] and by Choi et al [20]. Though there is appreciable scatter in the data, clearly $\Delta^{norm}$ increases with decreasing disorder. A linear fit gives an energy gap for vanishing disorder of $\Delta_{5/2}$ ~ 0.006 – 0.007 $e^2/\varepsilon l_B$.

A coefficient of 0.006 – 0.007 is within a factor of 2 – 3 of the most recent theoretical calculation of $\Delta_{5/2}/e^2/\varepsilon l_B$ ~ 0.016 [43]. Considering that the calculation [43] was carried out employing the parameters of the sample we are presenting here, and that the finite thickness of the 2DES and Landau level mixing effects [44,45] had been taken into account, this remaining difference of a factor of 2-3 seems to suggest that there exists an interplay between disorder and electron-electron interaction that goes beyond a simple level broadening. In this regard, we note that experimental energy gaps can differ significantly between very similar sample structures. For example, the sample structure, electron density and mobility are very similar in Ref. [17] and in this work and yet the gap data differ by 50% (Ref. [17] 0.3K, this work 0.45K). This large difference suggests that mobility is not the best measure to quantify disorder in these samples and the extrapolation in Fig. 1b is not reliable. Other sample parameters, such as the larger scale distribution of disorder distribution, may also play a significant role.

Alternatively, the 5/2 state may not be of the Pfaffian type. In fact, a so-called "anti-Pfaffian" state was recently proposed as an alternative candidate for the ground state at $\nu$=5/2 [46,47]. An energy gap has not been calculated yet for this state, but it may turn out to be closer to experiment, providing some hint as to the true correlation at 5/2.

### 3.2  Even-denominator FQHE state at $\nu$=19/8

The observation of a new even-denominator FQHE state at $\nu$=19/8 was first reported in Ref [18]. This state is very fragile, occurs only in very high quality samples and at very low temperatures. No other observation of the $\nu$=19/8 state has been made. Therefore, its parameters in the present specimen are important to report.

Fig. 2a shows the value of $R_{xx}$ at the minimum of the $\nu$=19/8 state as a function of $1/T$. The data show considerable scatter at higher temperatures ($T$ higher than ~20mK), and develop an activated behavior at lower temperatures. The scatter of the data and the limited range of $R_{xx}$ variation reflect the fragility of the state. A linear fitting at low temperatures (see Fig. 2a) yields an energy scale of ~ 5mK. Considering the limited temperature range and small change in $R_{xx}$, the obtained ~ 5mK most likely is not the true energy gap at $\nu$=19/8. Yet higher quality specimens and lower temperatures seem to be required to address this shortcoming.

To quantify the development of the state at $\nu$=19/8 and convince ourselves further that it represents a FQHE, we compare in Fig.2b the derivative of $R_{xy}$ at $\nu$=19/8 and at $\nu$=12/5 as a function of T. The data are reproduced from Ref. [18], with the addition of our most recent data point at $T$ ~ 6 mK. Both fractions show very similar behavior, moving from their classical high temperature $dR_{xy}/dB$ value towards the vanishing slope of a quantum Hall plateau at low temperatures. The $\nu$=12/5 state reaches this vanishing value, whereas the $\nu$=19/8 state falls slight short. Extrapolating towards $dR_{xy}/dB$=0 it appears that a temperature of ~2-3mK is required to reach a flat Hall plateau, which is consistent with the ~ 5 mK energy scale obtained from the $T$-dependence of $R_{xx}$ minimum in Fig 2a.

At present, the origin of the $\nu$=19/8 FQHE state is unknown. We speculate that it may also be a paired CF state, similar to the state at $\nu$=5/2 state. If this were the case, the mental sequence of creating the $\nu$=19/8 (=2+3/8) state would be to first map the partially filled 3/8 state onto the $\nu$*=3/2 state of CFs with two attached flux quanta (or $^2$CFs), where $\nu$* is the effective filling factor of $^2$CFs. Then, two additional flux quanta are attached to the $^2$CFs in the top, half-filled CF Landau level, thus, transforming the $^2$CFs to $^4$CFs. Ultimately pairing of $^4$CFs would give rise to the FQHE at $\nu$=19/8. Following this rationale, FQHE states may also exist at $\nu$=9/4 or 11/4. However, as will be shown in the following section, the $\nu$=9/4 and 11/4 states are composite fermion Fermi sea state. The reason for a different behavior at $\nu$=19/8 and $\nu$=9/4 or 11/4 may result from fact that at $\nu$=19/8, there is one fully-filled CF Landau level beneath, whereas at $\nu$=9/4 and 11/4 no such fully-filled CF Landau level exists. Consequently, the 19/8 CF state reflects more closely the 5/2 electron state, which is a FQHE, whereas the $\nu$=9/4 and 11/4 CF states reflects the 1/2 electron state, which is a CF Fermi sea state. As a final note we add that a FQHE state has been observed recently at $\nu$=3/8 in the lowest Landau [13]. Several proposals as to the origin of this state have been put forward [48-51], including p-wave pairing of CFs in the spin reversed sector [48] and clustering of composite bosons [51]. It must be left to future experiments to determine whether there is a connection between $\nu$=3/8 and $\nu$=19/8.

3.3     Odd-denominator FQHE states in the N=1 Landau level

The origin and the stability of most of the odd-denominator FQHE states in the N=1 Landau level remains a largely unresolved issue [33-39]. Compared to the lowest Landau level, much fewer odd-denominator FQHE states are observed in this Landau level. In fact, to date, only four, at $\nu$=7/3, 8/3, 12/5 and 14/5, are firmly established. The states at $\nu$=14/5 and 11/5 are generally believed to be of the Laughlin type [33]. The origin of the states at $\nu$=7/3 and 8/3, on the other hand, remains unclear. Earlier on, a Laughlin type FQHE state was ruled out for these states based on small, finite size, few particles calculations [33,52]. Later calculations, with larger numbers of particles seem to allow for a Laughlin type sate at these filling factors [34]. This particle number dependence differs from the stability of the 1/3 state in the lowest Landau level, which is the original Laughlin state and shows incompressibility at all sizes of systems [53].

In the $\nu$=12/5 state a so-called parafermionic state [36] might be realized. Yet it remains puzzling why there is no signature of a FQHE state at $\nu$=13/5, the particle-hole conjugate state of the $\nu$=12/5 state. In this section, we will present temperature dependent data at $\nu$=7/3 and 8/3 and discuss the absence of the $\nu$=13/5 state.

Fig.3 shows $R_{xx}$ at $\nu$=7/3 and 8/3 as a function of 1/$T$. The derived energy gaps are $\Delta_{7/3}$=0.59K and $\Delta_{8/3}$=0.29K. The relationship of $\Delta_{7/3}$ ~ 2 x $\Delta_{8/3}$ is unexpected, although this ratio has now been observed in two samples of very different electron density and mobility [16]. Theoretically, on the other hand, these two states are treated as electron hole mirrors and their energy gap is therefore expected to be the same at the same B field (which roughly holds, since (8/3)/(7/3) ~1). Recent experiment performed by others on similarly high quality samples, indeed, show activation energy gaps that are similar for 7/3 and 8/3 [20]. One needs to await further, low temperature data to reexamine the gap at the thirds in the first excited Landau level.

Signs of developing FQHE states are also observed at $\nu$=25/9 and, at higher temperatures, at $\nu$=19/7 (as shown in Fig.4). The sequences around $\nu$=11/4 (2+3/4), from 14/5 (2+4/5) to 25/9 (2+7/9) on the lower magnetic field (*B*) field side, and from $\nu$=8/3 (2+2/3) to $\nu$=19/7 (2+5/7) on the higher *B* field side, resembles those in the lowest Landau level around $\nu$=3/4. From this observation, together with an un-quantized transport behavior at $\nu$=11/4, one may conclude that the state at $\nu$=11/4 is also a CF Fermi sea state. The same probably holds at its electron-hole symmetric filling factor of $\nu$=9/4.

Surprisingly, in this ultra high quality sample, the $\nu$=13/5 state is totally missing, while its particle-hole conjugate state at $\nu$=12/5 shows a fully developed FQHE. This is not universally observed, since in previous data both fractions showed comparable strength [16]. To emphasize this absence of 13/5 in the present data, we show in Fig. 4 $R_{xx}$ at two temperatures, $T$ ~ 6 and 36 mK. At $T$ ~ 6 mK, for the $\nu$=12/5 state, $R_{xx}$ is very small and ~ 5 ohm and $R_{xy}$ (not shown here) is precisely quantized to better than 0.02%, using $R_{xy}$ at $\nu$=5/2 as a reference. At $T$ = 36mK $R_{xx}$ rise to ~ 56 ohm and exhibits thermally activated behavior in between (not shown). At $\nu$=13/5, on the other hand, there is no evidence of a FQHE state at neither temperature.

The slight difference in *B*-field at $\nu$=12/5 and $\nu$=13/5 hardly can explain the absence of the 13/5 state, given that the activation energy at $\nu$=12/5 is as large as 70mK [18]. Alternatively, the $\nu$=13/5 state may be affected by the existence of the neighboring re-entrant integer quantum Hall state (RIQHE) at $\nu$~2.56. However, signs of a weaker FQHE state usually are observable at higher temperatures when the earlier overpowering state subsides in strength. This is not the case here, as can be seen in Fig. 4, at *T* ~ 36 mK, where, in spite of the weakness of the RIQHE, there is no sign of a $\nu$=13/5 FQHE state, while $\nu$=12/5 is well developed. Towards a third explanation, we recall the existence of a $\nu$=13/5 state, as strong as the $\nu$=12/5 state, in previous experiment [16], in a specimen of smaller electron density and thus at smaller *B* field, which favors spin flips. The absence or presence of a $\nu$=13/5 state may therefore be spin related , with a transition from a spin unpolarized (or partially polarized) at small *B* fields to  a spin-polarized state around *B* ~ 5T. Of course, there are other effects that may lead to the breaking of particle-hole symmetry and thus to a disappearance of the $\nu$=13/5 state, such as Landau leveling mixing, or finite thickness [54]. Indeed, the sample of Ref. [16] is a single heterojunction while the sample of and Ref. [18] is a quantum well and this may affect which of these weak FQHE states can be observed.

3.4    Temperature dependence of the reentrant integer quantum Hall effect states around $\nu$=5/2

In the N=1 Landau level between $\nu$=2 and 3, aside from the FQHE states, another distinguishing feature is the so-called reentrant IQHE state at  Landau level fillings $\nu$~2.30, 2.44, 2.56, and 2.70. At these values, $R_{xx}$ becomes vanishingly small at very low temperatures and $R_{xy}$ abruptly changes from the classical Hall value and assume a quantized plateau with the value of the closest IQHE state. Though its origin is still unresolved, the connection between this reentrant phase and the bubble phases in the third and higher Landau levels [55-57] have been suggested [58,59,51]. Recently, it was shown that these RIQHE states are very sensitive to an in-plane magnetic field and disappear very quickly within a few degrees of tilt of the *B*-field [19]. This transport behavior was interpreted as a tilt-induced melting of the bubble phase [19]. To complete our study of the first excited Landau level presented in this paper we add here some data on the melting behavior, by raising the temperature, of the two RIQHE states around $\nu=5/2$.

Fig. 5 shows the values of $R_{xy}$ for the RIQHE versus temperature. Both show a very similar behavior: $R_{xy}$ remains quantized to an integer value at very low temperatures, increase very quickly within a small temperature range, and assume their respective classical values thereafter. The temperature range where $R_{xy}$ varies quickly is different for these two RIQHE states. For the $\nu\sim2.44$ state on the higher *B* field side, it ranges from 25 to 35 mK, while for the state at $\nu\sim2.56$ on the lower *B* field side ranges from 35 mk to 50 mK. Over the same temperature range the 5/2 state remains a good quantum Hall state.

This abrupt *T* dependence of $R_{xy}$ in the RIQHE has been observed earlier [19], and was attributed to the melting of an assumed, underlying bubble phase whose energy scale is expected to scale as $e^2/\varepsilon l_B$. Therefore, invoking again electron hole symmetry, one would expect the RIQHE at $\nu\sim2.44$ to be more stable than the RIQHE state at $\nu\sim2.56$ Yet, we observe exactly the opposite. Here too, we have no basis to explain the apparent contradiction, but bring up the recently promoted Pfaffian and anti-Pfaffian phases [46,47] that might play a role, both of which break electron-hole symmetry and affect features around $\nu=5/2$ depending whether they occur on the electron or hole side of 5/2.

4. Summaries and open questions

In an ultra-high mobility two-dimensional electron system, at ultra-low temperatures, we observe a very complex electronic transport behavior in the first excited Landau level. In detail, the $\nu=5/2$ state is very strong in this specimen and its energy gap is 0.45K. Residual disorder has an important impact on the value of the energy gap at $\nu=5/2$. For another even-denominator FQHE state at $\nu=19/8$ an energy scale of ~5mK is deduced for its gap. This small energy scale attests to the need of still lower electron temperature and/or high sample quality for the $\nu=19/8$ FQHE state to be fully developed. As for the odd-denominator FQHE state, we measured the energy gap at $\nu=7/3$ and 8/3. Like the 5/2 state, their energy gaps increase with decreasing disorder. The state at $\nu=12/5$ is developed into a full FQHE state, however, its particle-hole conjugate state, the $\nu=13/5$ state, is entirely missing in this ultra-high quality 2DES. We speculate that this disappearance might be related to the spin polarization of the $\nu=13/5$ state. For the two-reentrant integer quantum Hall states around $\nu=5/2$ we observe a temperature scale that is opposite to the expected behavior, which is puzzling and may be related to the recently proposed Pfaffian and anti-Pfaffian states.

Of all the electronic states in the first excited Landau level, the state at $\nu=5/2$ remains the most exciting, but also quite enigmatic. Based on the p-wave pairing within the CFs model, the $\nu=5/2$ state is expected to be spin polarized. Even this, relatively simple property is not totally substantiated experimentally. So far, we only have indirect evidence of a spin-polarized 5/2 state from a tilted *B* field induced anisotropy [30,60,61]. The density dependence of the $\nu=5/2$ gap could shine light on the spin polarization, as with increasing density and hence increasing *B*, a collapse of the gap due to a spin transition may be observed or be absent. A previous such measurement [62] did not observe a spin transition and hence favored the existence of a spin polarized state in support of the CF pairing model and the Pfaffian state. While this experiment employed a relatively low quality heterojunction insulated gate field effect transistor (HIGFET), in which the $\nu=5/2$ state was not fully developed, it would be desirable to revisit this question in a future, high quality HIGFET. The resistively detected NMR technique has proven to be a very powerful tool in directly measuring the spin polarization of FQHE states [63]. This same technique has been contemplated for studying the 5/2 spin state. However, RF heating of the specimen has so far prevented any conclusion as to the spin. While the Pfaffian state remains the front runner in explaining the existence of the state at $\nu=5/2$, an anti-Pfaffian state was proposed recently as an alternative candidate [46,47]. Yet their spin polarization is not a tool to discriminate between both.

The spin polarization of the 7/3 and 8/3 states in the N=1 Landau level so far is largely unpursued. Naively, extrapolating from the lowest Landau level, one might expect that the 7/3 state is spin polarized, whereas the 8/3 state is unpolarized. However, at least one theoretical paper [64] predicts that, contrary to our intuition, the $\nu$=8/3 state is also spin-polarized. Experimentally, the recent study at ultra-low temperature, showed a surprisingly complex tilted magnetic field dependence of the 7/3 and 8/3 states [19]. These third states in the N=1 Landau level may be much more complex than expected.

The disappearance of the $\nu$=13/5 state continues to be puzzling. Whether its disappearance is a result of a spin transition at the particular *B*-field of our present experiment is unclear here too. Spin may be a primary ingredient.

The exciting new even denominator state at $\nu$=19/8 needs further confirmation. At this point it is still shy of showing the ultimate, characteristics of a true FQHE state. To this end, higher sample quality and lower electron temperature are needed.

In general, it appears that spin may be the essential ingredient for the behavior of many states in the N=1 Landau level. The occurrence of the related features at typically lower *B*-field than the equivalent states in the lowest Landau level makes such a conjecture quite likely. Imaginative, new experimental techniques as well as yet higher quality specimens seem to be required to further assess electron-electron correlation in this first excited Landau level.

## 5. Acknowledgment

We thank Th. Jolicoeur for helpful discussions. This work was supported, in part, by DOE/Basic Energy Science. Sandia is a multiprogram laboratory operated by Sandia Corporation, a Lockheed Martin company, for the United States Department of Energy's National Nuclear Security Administration under contract DE-AC04-94AL85000. The work at Columbia was supported by DOE and W.M. Keck Foundation. The work at Princeton was supported by the AFOSR, the DOE, and the NSF. Experiment was performed at the high *B/T* facilities of the National High Magnetic Field Laboratory, which is supported by NSF Cooperative Agreement No. DMR-9527035 and by the State of Florida.

Table I: List of FQHE states discovered to date. States with (?) have been observed as particular features in $R_{xx}$ and/or $R_{xy}$, but the accuracy of their quantization has not been established.

| 1/3 | 1/5 | 1/7 | 1/9 | 2/11 | 2/13 | 2/15 | 2/17 | 3/19 | 5/21 | 6/23 | 6/25 |
|---|---|---|---|---|---|---|---|---|---|---|---|
| 2/3 | 2/5 | 2/7 | 2/9 | 3/11 | 3/13 | 4/15 | 3/17 | *4/19* | 10/21 | | |
| 4/3 | 3/5 | 3/7 | 4/9 | *4/11* | *4/13* | 7/15 | 4/17 | 5/19 | | | |
| 5/3 | 4/5 | 4/7 | 5/9 | 5/11 | *5/13* | 8/15 | *5/17* | 9/19 | | | |
| 7/3 | 6/5 | 5/7 | 7/9 | 6/11 | 6/13 | 11/15 | *6/17* | 10/19 | | | |
| 8/3 | 7/5 | 9/7 | 11/9 | *7/11* | 7/13 | 22/15 | 8/17 | | | | |
| | 8/5 | 10/7 | 13/9 | 8/11 | 10/13 | 23/15 | 9/17 | | | | |
| | 9/5 | 11/7 | 14/9 | 14/11 | 19/13 | | | | | | |
| | 11/5 | 12/7 | 25/9 | 16/11 | 20/13 | | | | | | |
| | 12/5 | 16/7 | | 17/11 | | | | | | | |
| | 13/5(?) | 19/7 | | | | | | | | | *5/2* |
| | 14/5 | | | | | | | | | | *7/2* |
| | 16/5 | | | | | | | | | | *3/8(?)* |
| | 19/5 | | | | | | | | | | *5/8(?)* |
| | **21/5** | | | | | | | | | | *19/8* |
| | **24/5** | | | | | | | | | | *3/10(?)* |

Table II: Parameters of five ultra-high mobility specimens -- density, mobility, and energy gap at $\nu=5/2$.

| Sample number | Density ($10^{11}$ cm$^{-2}$) | Mobility ($10^6$ cm$^2$/Vs) | Energy gap at $\nu=5/2$ (K) |
|---|---|---|---|
| A | 3.1 | 31 | 0.45 |
| B | 3.2 | 28 | 0.22 |
| C | 2.3 | 26 | 0.24 |
| D | 3.0 | 20 | 0.26 |
| E | 2.2 | 17 | 0.11 |

Figures:

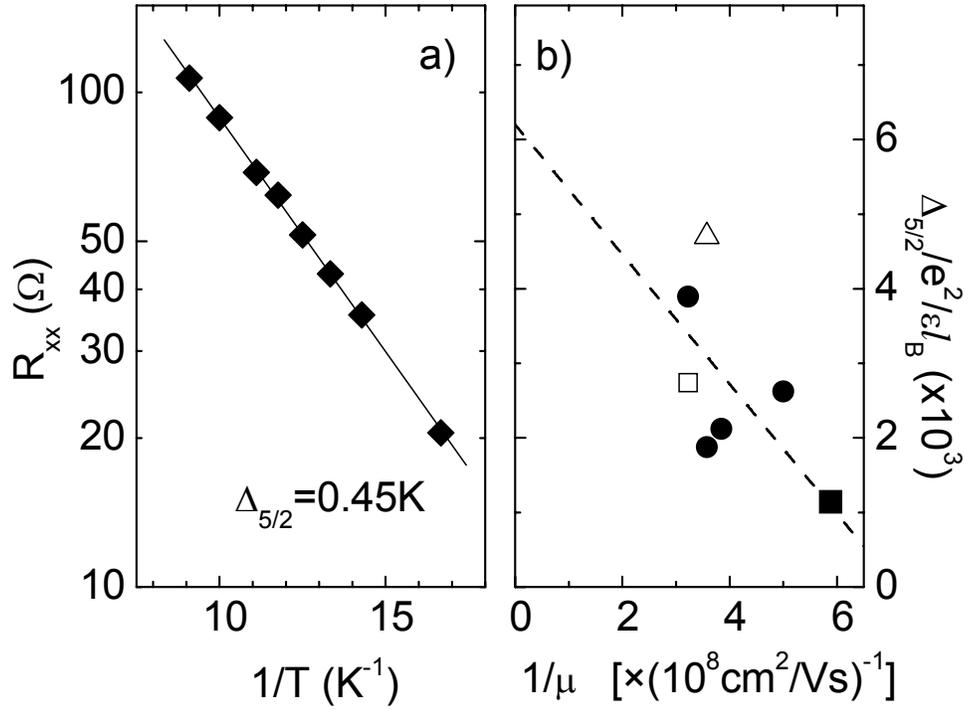

Figure 1: a) Arrhenius plot for the $R_{xx}$ minimum at $\nu=5/2$. The line is a linear fit. b) Normalized energy gap $\Delta^{norm} = \Delta_{5/2}/e^2/\varepsilon l_B$ for five samples of different mobilities. Results from Ref. 17 (open square) and Ref. 20 (open triangle) are included. The line shows a linear fit to the data points.

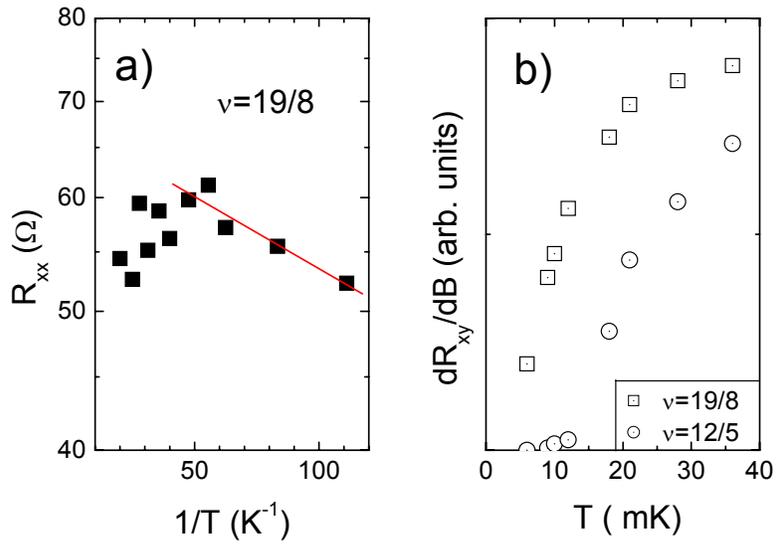

Figure 2 a) Temperature dependence of the $R_{xx}$ minimum at $\nu=19/8$. The line is a linear fit. b) Temperature dependence of the derivative of the Hall resistance $R_{xy}$ at $\nu=19/8$ and 12/5. This plot is reproduced from Ref. [18], with an extra data point at $T = 6$ mK.

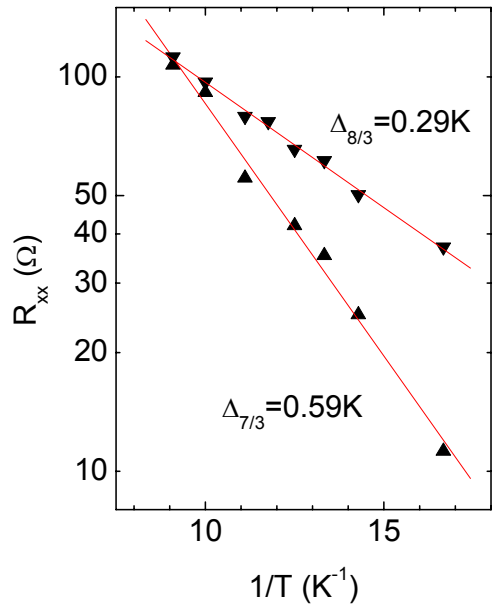

Figure 3: Arrhenius plot for the $R_{xx}$ minima at $\nu=8/3$ and 7/3. Lines are linear fits.

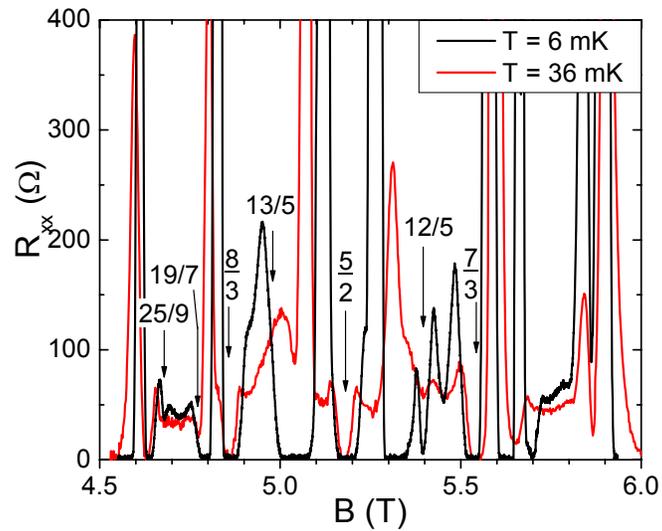

Figure 4: $R_{xx}$ between $\nu=2$ and 3 at two temperatures $T \sim$ 6 and 36 mK. Landau level fillings are marked by arrows.

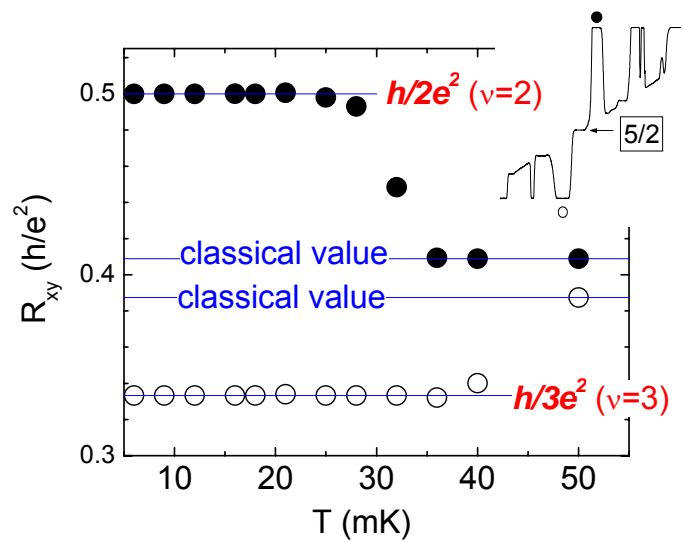

Figure 5: Temperature dependence of $R_{xy}$ of the two RIQHE states around $\nu=5/2$. The horizontal lines mark the values of nearby integer Hall plateaus and the corresponding classical Hall values.